\documentclass[a4paper, 12pt, oneside]{article}

\usepackage[utf8]{inputenc}
\usepackage[english]{babel}
\usepackage{graphicx}
\usepackage[margin=1in]{geometry}
\usepackage{rotating}
\usepackage[super,comma]{natbib}
\usepackage{caption,tabularx}
\usepackage{latexsym}
\usepackage{slashbox}
\usepackage{multirow}
\usepackage{authblk}
\usepackage{bbm}
\usepackage{comment}
\usepackage{dcolumn}
\usepackage{url}
\usepackage{hyperref}
\usepackage{booktabs}
\usepackage{array}
\usepackage{adjustbox}
\usepackage[usenames,dvipsnames,svgnames,table]{xcolor}
\usepackage[toc,page]{appendix}
\usepackage[linesnumbered,ruled]{algorithm2e}
\usepackage{tikz}
\usetikzlibrary{shapes.geometric, arrows, calc, fit, positioning, chains, arrows.meta}

\usepackage{subcaption}
\usepackage[linesnumbered,ruled]{algorithm2e}
\usepackage{multicol}

\usepackage{booktabs}

\usepackage{amsmath, amsthm, amsfonts}
\usepackage{amssymb}
\usepackage{float}

    \usepackage{newfloat}
    \DeclareFloatingEnvironment[name={Extended Data Fig.}]{suppfigure}

{
  \title{\bf  
  A Unique Cardiac Electrophysiological 3D Model}
 }

\author[1,2,*]{Cristina Rueda}
\author[1]{Alejandro Rodr\'iguez-Collado}
\author[1,3]{Itziar Fern\'andez}
\author[1]{Christian Canedo}
\author[4]{Mar\'ia Dolores Ugarte}
\author[1,2]{Yolanda Larriba}

\date{}

\affil[1]{Department of Statistics and Operations Research,
University of Valladolid, Spain}
\affil[2]{Mathematics Research Institute of the University of Valladolid (IMUVA), Spain}
\affil[3]{Biomedical Research Networking Center in Bioengineering, Biomaterials and Nanomedicine (CIBER-BBN),  Spain}
\affil[4]{Healthcare Research Institute of Navarre (IdiSNA),   Spain}
\theoremstyle{definition}
   
   \newtheorem{Def}{Definition}

\begin{document}
\maketitle

\begin{abstract}
 Mathematical models of cardiac electrical activity are one of the most important tools  for elucidating information about the heart diagnostic. Even though it is one of the major problems in biomedical research, an efficient mathematical formulation for this modelling has still not been found.
  In this paper, we present an outstanding mathematical model. It relies on a  five dipole representation of the cardiac electric source, each one associated with the well-known waves of the electrocardiogram signal. The mathematical formulation is simple enough to be easily parametrized and rich enough to provide realistic signals. Beyond the physical basis of the model, the parameters are physiologically interpretable as they characterize the wave shape, similar to what a physician would look for in signals, thus making them very useful in diagnosis.
  The model accurately reproduces the electrocardiogram and vectocardiogram signals of any diseased or healthy heart, bringing together different systems in a single model. 
  Furthermore, a novel algorithm  accurately identifies the model parameters.
This new discovery represents a revolution in electrocardiography research, solving one of the main problems in this field. It is especially useful for the automatic diagnosis of cardiovascular diseases, patient follow-up or decision-making on new therapies.

\end{abstract}

\section{Main}
The development of models and algorithms for the study
of the cardiac electric system helps to better understand the physical processes governing the  system  and  helps to guide therapeutic planning.  The relevance of the topic has attracted the interest of scientists from different fields, such as mathematics, physics, bioengineering and medicine.
\\

The heart muscle is a composite tissue with a complex structure that
consists of various cell types. 
The electric activation of the heart begins at the sinus node, where pacemarker cells  activate spontaneously. The corresponding current results in the excitation of the neighbouring cells, and it then  spreads, first along the atria, and then along the ventricles\cite{Mal95,Bay07}.

Generalised assumptions on cardiac electrophysiological models that date back to the 1960s (ref.\cite{Hol69}), and remain part of the conventional approach are: the electric field is represented by a single or multiple dipoles, while the total electric activity is represented by a three-dimensional vector. This vector, denoted by $\vec{D}(t)$ at time $t$, is the sum of all the individual dipole vectors. As the depolarization wavefront spreads through the heart,  $\vec{D}(t)$ changes in magnitude and direction as a function of time.  $\vec{D}(t)$  typically describes a trajectory with three loops corresponding to consecutive time segments: the $P$ wave (atrial depolarization), the $QRS$ complex (ventricular depolarization) and the $T$ wave (ventricular repolarization), respectively. The loops described by the $P$ and $T$ waves are elliptical, while the $QRS$ has an irregular shape \cite{Jar19}. Furthermore, the voltage measurements registered by the Electrocardiogram (ECG) signal are the projections of $\vec{D}(t)$ in the directions of the axes of the recording electrodes located on the thoracic surface. 
In general, positive (negative) signals are produced when the depolarization front propagates  towards (away from) a positive electrode. The opposite happens for the repolarization front. The standard ECG  has signals from 12 projections or leads recorded using 10 electrodes \cite{Bay07}.
 \\

In spite of  the previous assumptions,  mathematical  formulations for ECG signals remain quite complex. Classical formulations  
include differential equation systems, representing the process with more or less  biophysical detail. 
 Aside from the mathematical complexity of the model formulation,  some common criticisms of these models are  that they hardly generate realistic 12-lead ECG signals,  and that they depend on a large number of parameters which are hardly identifiable. In practice, a meaningful parameter identification is essential.

 The literature dealing with dipole models for the  electric activity of the heart is very extensive\cite{Cla11, Bal15, Elh17, Qua17, Nie19, Qui19, Whi20, Ver20, Cor21, Das21, Che21}. 
 Alternative approaches dealing with specific aspects of the heart's electric activity are also too many to be easily summarised\cite{Sch17, Nay18, Sat18, Han19, Han20, Sio21}.
 Furthermore, many papers since the early 1950s have been devoted to Vectorcardiography (VCG) models\cite{Jar19}. Instead of 12-lead standard ECG models,  they only consider three leads scanned in quasi-orthogonal axes. However, the VCG system is not common in clinical practice as there are fewer experts trained in these signals.
ECG and VCG models have contributed to an improved understanding of the functioning of the heart, but have so far had little success in convincing clinicians. In particular, they fail in an important prerequisite for clinical applications,  the ability to faithfully replicate ECG recorded from any diseased or healthy heart.
\\

\begin{figure}
\centering
				\includegraphics[width=0.75\textwidth]{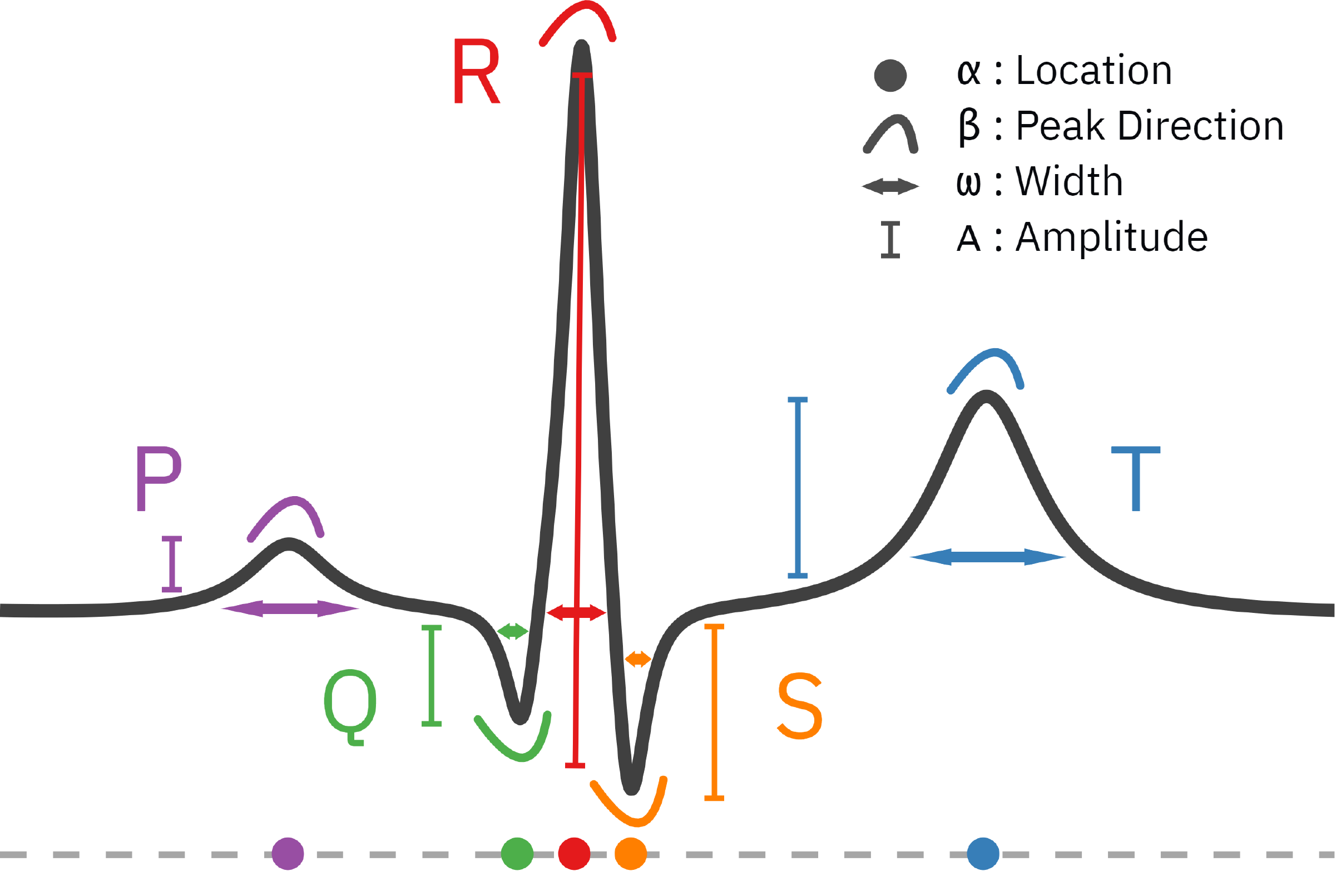}
				\caption{FMM parameter description. An ECG healthy heartbeat with waves decomposition and morphological parameter description.}
				\label{f:fi0}
			\end{figure}

Here we present a novel model, named 3DFMM$_{ecg}$ that relies on the classical physiological premise: the electric source is represented by a multiple dipole model. The novelty, however, is that it assumes that  $\vec{D}(t)$ combines the electric signals from exactly five different sources, which represent differentiated myocardium segments,  further associated to the five fundamental waves in ECG signals. Namely, $\vec{D}(t)=\vec{d}_P(t)+\vec{d}_Q(t)+\vec{d}_{R}(t)+\vec{d}_S(t)+\vec{d}_T(t)$, where $t$ represents the real time and  varies in $(0,2\pi]$ for each heartbeat. Furthermore, the new  model is also based on a key idea and original assumption:   $\vec{d}_J(t)$, with $J \in \{P,Q,R,S,T\}$, moves in the same plane as time progresses, and its trajectory is described as a complex FMM signal, a  parametrized mathematical equation that describes an elliptical trajectory. Accordingly, the magnitude of $\vec{d}_J(t)$ in a given direction is  represented by a one-dimensional FMM wave, which reflects depolarization and  repolarization voltage changes.  An FMM wave, $W(t,A,\alpha,\beta,\omega)$,  is an equation defined in terms of four parameters \cite{Rue19}. The parameter $A$, a positive real number, measures the wave amplitude;  $\alpha$ is a location parameter with values in $(0,2\pi]$; while $\beta$, with values  in $(0,2\pi]$, describes the asymmetry degree of the wave shape and also informs  when the unimodal pattern corresponds to a crest or a trough. Finally, the parameter $\omega$, with values in $[0,1]$, measures the sharpness of the peak. The value $\omega=1$ corresponds to an exact sinusoidal shape.  Figure \ref{f:fi0} illustrates each parameter in a simulated ECG signal.
These four parameters are wave-specific, but, while $A$ and $\beta$ are also lead-specific, $\alpha$ and $\omega$ are the same across leads. The role of these common parameters is essential in the modelling process. They provide  connectivity between the signals from different leads thus,  simplifying the model substantially. 
 For a given lead $L$ and wave $J$, the quadruple  [$A_J^L,\alpha_J, \beta_J^L,\omega_J$] describes precisely the wave shape, in the same way a physician looks at the signal. 
From the model definition it also follows that projections of $\vec{D}(t)$, in directions different from those in the standard 12-lead system, are also formulated as a sum of five FMM waves where $\alpha$ and $\omega$ do not change. Thus, the approach unifies VCG and ECG systems within a single formulation.

   The inclusion of an intercept and an error term that accounts for  data noise results in a powerful statistical model.
To make the model identifiable and even more biologically interpretable, we assume that $\alpha_P\le\alpha_Q\le\alpha_R\le\alpha_S\le\alpha_T$. These restrictions respond to the natural depolarization order  of different myocardium segments.  These restrictions and the fact that the number of parameters is small, some of them common to different leads, guarantee an accurate identification of the model parameters, even in the cases of highly anomalous patterns or noisy signals.   
A quadratic optimization problem must be solved to obtain the model parameters. Here, we design a backfitting iterative approach to solve the problem. Specifically, the common parameters, $\alpha$ and $\omega$, are identified using a grid search. Then, the lead-specific parameter estimates $A$ and $\beta$ are easily derived using standard linear regression models. 
  An algorithm is designed to analyse 12-lead ECG fragments of any length. In the preprocessing stage, the signal is divided beat by beat. Then, for each beat, parameter estimates are derived.
  The output of the algorithm provides, for each wave, the series of parameter values, corresponding to consecutive beats, which can be summarised to get  average patterns  as well as the changes in the patterns over time. 
\\

\begin{figure} 
				\includegraphics[width=1\textwidth]{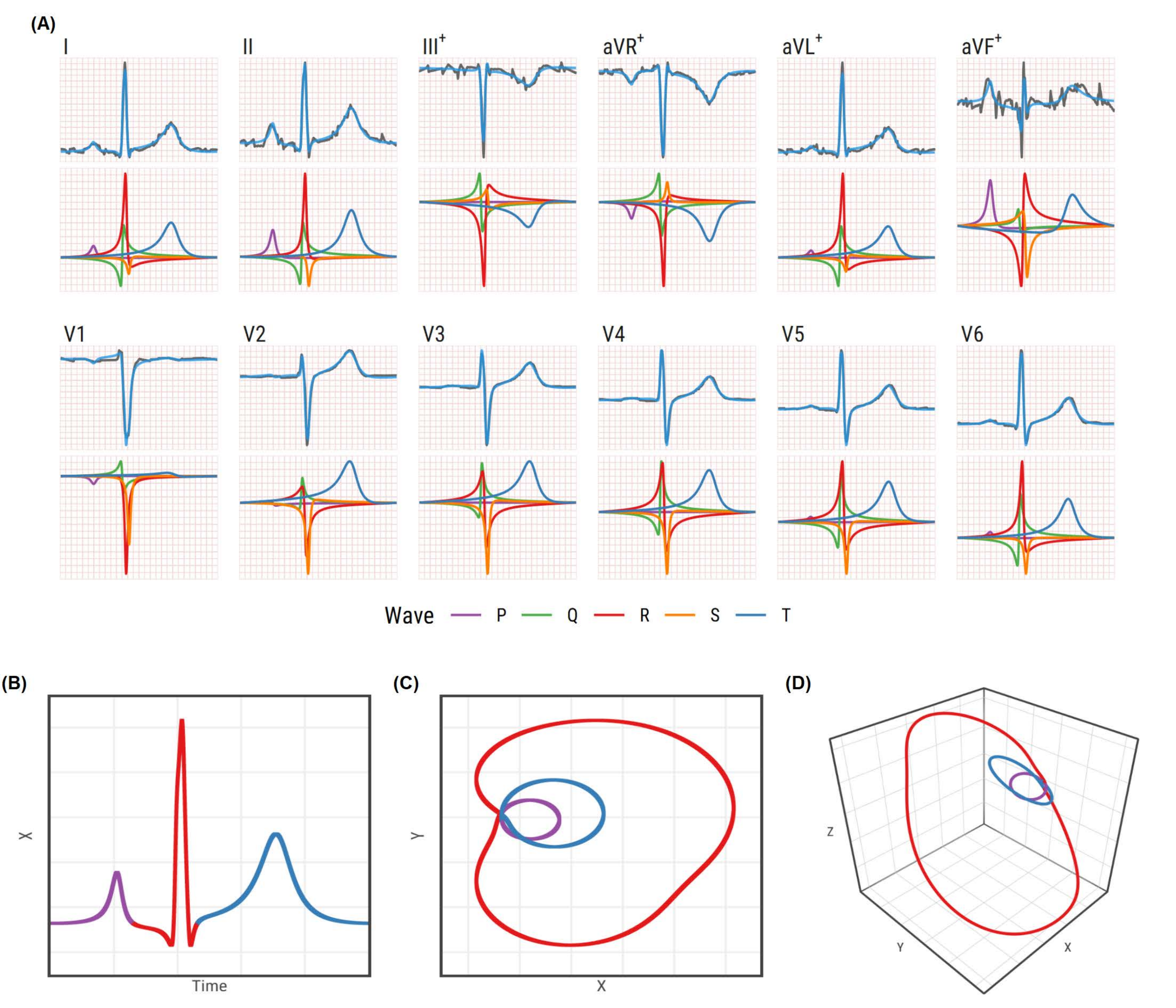}
				\caption {ECG for patient 8 in the PTB-XL database, beat nº9. 
				(A) 12-lead signals, using individual voltage scales (not shown).  Observed (grey) and FMM predicted (blue) signals  in the top panels, and wave patterns in the bottom panels. The symbol + indicates linear combinations of other signals.  (B), (C) and (D):  1D, 2D and 3D representations of the predicted signals, respectively, where $X= LeadII; Y= LeadII^{(i)}; Z= LeadV2-2*Y$. The colours differentiate the $P$, $QRS$ and $T$ loops. }
				\label{f:figure1}
			\end{figure}

To evaluate the performance of the 3DFMM$_{ecg}$ model, we proceed as follows. Firstly, we analysed the model's ability to reconstruct realistic  ECG signals, and hence solve the forward problem, using  1D, 2D and 3D representations generated from different parameter configurations. Secondly, all the  ECG fragments from patients in the PTB-XL database \cite{Wag20} were analysed with the new model. This database  contains a great variety  of anomalous, pathological and noisy cases. PTB-XL is a benchmark database, that has been used, in particular, for the challenge of Computing in Cardiology 2020 (ref. \cite{Per20}). The quality of model reconstruction is very high for the 12-lead ECG signals in most cases, something that other models are very far from reaching. Furthermore, different pathologies are associated with different parametric configurations, which indicates  how the model also solves the inverse problem. Specifically, three new ECG markers are defined in terms of the model parameters that are useful in the diagnosis.  Figure \ref{f:figure1} illustrates how the model works in a healthy heart. The choice of axes for the  2D and 3D representations is explained later on.
Finally, we have designed an app  (https://fmmmodel.shinyapps.io/fmmEcg3D/) to make it easier for researchers to check the model with their own data. 
\\

Next sections show how the 3DFMM$_{ecg}$ successfully solves the main challenges of the forward  and  inverse problems in electrocardiography. 
  Furthermore, a simulation study   that shows how  the identification algorithm provides sensible and robust parameter estimates across different configuration scenarios was conducted (see Supplementary Information). The analysis of real ECG data requires a preprocessing stage to remove baseline and other noise artifacts. Data preprocessing is detailed in the Supplementary Information.

\section*{Reconstruction of realistic ECG signals}
A configuration of parameters for a healthy heart, labelled NORM, has been identified using the median values for all patients, labelled as NORM in the PTB-XL database. This configuration is used as a reference. Alternative configurations describing different pathological conditions of the heart have been generated by changing the reference values of the most relevant parameters, summarised in Table S1 of the Suppementary Information.
The standard 12-lead unidimensional signals, and selected 2D and 3D representations, are provided for each configuration scenario. The axes we have selected for the 2D and 3D representations are defined as follows; 

\begin{equation}
 X= LeadII; Y= LeadII^{(i)}; Z= LeadV2-2*Y \nonumber
\end{equation}

$X$ is the most commonly used lead in studies, and $(X,Y)$ is a natural bidimensional representation described by the Analytic Signal, defined in the complex plane  \cite{San18, Rue21a} (additional details are given in the Methods Section). Finally, $Z$ has been selected  after checking other alternatives, as it offers an interesting visualisation of the three loops: $P$, $QRS$ and $T$ across different patterns.

 To conserve space, the figures for these configurations are shown from the Extended Data Fig. \ref{fig:simNORM} to \ref{fig:simLVH}. In particular, the figure for the NORM configuration looks similar to Figure \ref{f:figure1}. The other configurations generate representations with characteristic and differentiated patterns.

\section*{Analysis of ECG signals from the  PTB-XL database}
In the preprocessing step, 210 patients (less than 1\%) out of 21,837, were discarded due to very noisy patterns. Additionally, 295 patients with pacemakers  were discarded due to their highly atypical but predictable ECG patterns.
 
A diagnostic label has been assigned to each patient in the database using SNOMED CT ontology \cite{Per20}, (\url{bioportal.bioontology.org/ontologies/SNOMEDCT}). Labels are assigned only to patients  with a diagnostic likelihood of 80 or more. Some of the more recognised SNOMED labels used here are Complete Bundle Branch Blocks  (CLBBB, CRBBB) left and right, respectively,  and ventricular Hypertrophy (HYP). 

We have calculated the model parameters for a given patient and beat. Furthermore, interesting summary measures are derived at patient level. Specifically, we define $omeR$ and $omeS$  as the median of the $\omega_R$ and $\omega_S$ individual beat values, respectively; and $maxAR$ is the maximum, across leads, of the median of the $A_R$ individual beat values.  In addition, a global model quality measure, named $\overline{R}$, is also obtained. It is defined as the mean, across leads, of an  $R-squared$ measure, which is the proportion of variation in the signal that is explained by the model and is a common evaluation metric in regression models.
\begin{table}
\small {
\begin{tabular}{lc|ccc|ccc|ccc|ccc} \toprule

Diagnostic & &\multicolumn{3}{c}{${\overline{R}}$} & \multicolumn{3}{c}{\textit{omeR}}& \multicolumn{3}{c}{\textit{omeS}}& \multicolumn{3}{c}{\textit{maxAR}} \\

& Nº &$p_{5}$&$p_{50}$&$p_{95}$&$p_5$&$p_{50}$&$p_{95}$&$p_5$&$p_{50}$&$p_{95}$&$p_5$&$p_{50}$&$p_{95}$ \\ \toprule

NORM & 8,578 & .91 & .95 & .97 & .03 & .03 & .04 & .02 & .03 & .05 & 525 & 837 &	1,379 \\ \midrule

CLBBB & 505 & .91 & .96 & .98 & \textbf{.07}&	\textbf{.14}&	\textbf{.21}&.04&	.06&	.10&806&	1,585&	2,673\\

CRBBB & 523 & .87 & .94 & .97 & .03 & .04 &	.11 & \textbf{.04} &	\textbf{.08}& \textbf{.14} & 396 & 811 &	1,627\\

HYP & 599 & .92& .96 & .98 & .03 & .04 & .09 & .02 & .03 & .06 &	\textbf{852} & \textbf{1,465} &\textbf{ 2,403}\\ 

\bottomrule \\[-0.75em]

ALL & 21,331 & .92 & .95 & .97& .03 & .04 & .09 & .02 & .03 & .07 & 520 & 888 & 1,677\\

\bottomrule

\end{tabular}
}

\caption{Statistics for selected parameters across diagnostics. Figures are a summary of the values for all patients which are, in turn, the median of the individual beat values.}
\label{t:STA}

\end{table}
 Table \ref{t:STA}  gives percentile ranges for these measures obtained across selected diagnostic classes. Specifically, normal ranges are provided.  The figures in Table \ref{t:STA} show  that the $\overline{R}$ values are  quite high across diagnostics, being higher than 90\% for more than 95\% of patients.  On the other hand,  Table \ref{t:STA} shows that $omeR$ and $omeS$ are good markers of CLBBB  and CRBBB, respectively, while $maxAR$ is a marker for HYP, whenever $omeR$ is within the normal range. 
The interest of the $omega$ FMM$_{ecg}$ parameters in the diagnosis had already been evidenced  in \cite[Rueda et al.][]{Rue21d}, where simple diagnostic rules for CLBBB and CBBB (Complete Bundle Branch Block) have been defined, and can be used even when  only one of the lead signals is available. Examples of observed and predicted signals from diverse heart conditions and noisy cases are given from the Extended Data Fig. \ref{fig:Ex1} to \ref{fig:Ex5}, illustrating the ability of the model to successfully handle anomalous patterns.

\section*{Discussion}

Here we present a unique cardiac electrophysiological model, the 3DFMM$_{ecg}$, which faithfully reconstructs the standard 12-lead ECG signal. The robustness of the method is assessed under a variety of both pathological and noisy conditions.
The model also succeeds in providing  reliable quantitative results to enhance the understanding of biophysical processes, including a functional connectivity between different lead signals. A relatively small set of parameters with morphological interpretation summarises the 12-lead signals, being  reliable for extrapolation and prediction, thus simplifying the diagnostic task. The model parameters are accurately estimated avoiding identifiability problems.  No other model had ever come close to achieving these goals. 
Furthermore, the results in the paper  validate the premise that the electrical source can be represented with a five two-dimensional dipole model, on which the new model is built. Interestingly, dipole models have been widely used in different forms in the literature. 
   Another great advantage of the model is its capacity for data compression,  as minimal hard disk space is needed to fully reconstruct the signal. The monitoring of patients and, in particular, telemedicine requires the storage of ECG signals and a fast on-line transmission of information \cite{Bhl19,Bha20}, and therefore efficient data compression methods are in high demand.
\\

Many challenges arise from this study. On the one hand, there is much work to do from a research perspective, such as the design of new diagnostic rules based on the FMM parameters; 
 the identification of wave patterns specific to exceptional conditions such as atrial flutter, which do not adapt so well to the model;  the integration of  the new parameters in deep or machine learning procedures for classification tasks \cite{Han19,Han20,Xue21,Sio21}; or the inclusion of new terms accounting for different sources of variability. Specifically, we will start by studying a rule  for HYP, using a combination of markers including the $maxAR$ defined in this paper. 
 On the other hand, there is  the committed objective of getting the new markers to be used in clinical practice.  
 To that end, devices that record the signals should provide the new markers, and more importantly, physicians need to be trained.
 Moreover, the current paper opens up great research  opportunities for studying biological electric systems beyond the cardiac one, which have an impact on the knowledge of the effect and causes of multiple diseases, as well as the development of drugs and therapies. In particular, electric signals from other organs, such as the brain or the eyes, can be modelled using adapted 3DFMM models. Thus, the important question of analysing the relationship between signals recorded in different regions, could also be addressed with these models. 
 Some work in this line is in progress.
\\ 

Regarding the limitations, the computational time is still too high  to give interactive outputs. This is due to the exhaustive search and the backfitting loop, which  are time-consuming processes. However, by implementing the current method in C and using general-purpose computing on graphics processing units, the computational cost could be greatly reduced. Furthermore, the algorithm, as described here, does not identify the $P$ wave when it is not located before the $QRS$ complex. New research into detecting and adapting the algorithm in these cases could bring about the correct identification of the $P$ wave.

\section*{Methods}

\subsection*{ 3DFMM$_{ecg}$ Model}
 
 Without loss of generality, it is assumed that the time points are in  $(0,2\pi]$. In any other case, transform the time points $t' \in [t_0,T+t_0]$ to $t=\frac{(t'-t_0)2\pi}{T}$.

Let us define an FMM wave as follows: $	W(t,A,\alpha,\beta,\omega)=A\cos(\phi(t,\alpha,\beta,\omega)),$ where $A$ is the wave amplitude and,
\begin{equation}\label{phi}
\phi(t,\alpha,\beta,\omega)=
\beta+2\arctan(\omega\tan(\frac{t-\alpha}{2})), t\in (0,2\pi]
\end{equation}
is the wave phase.

 $W(t,A,\alpha,\beta,\omega)$  is suitable for describing rhythmic up-down-up (or down-up-down) patterns, as is well justified in \cite[Rueda et al.][]{Rue19}. The parameters characterise various morphological aspects of the wave, as detailed in the introduction.
\\

An FMM complex signal, called Analytic Signal (AS), is defined using the Hilbert Transform (HT) as follows:
$S(t)=\mu(t)+i\nu(t)$; where,  $\mu(t)=W(t,A,\alpha,\beta,\omega)$, and $\nu(t)= HT(\mu(t))$.
This signal has interesting properties and researchers often assume that the underlying complex signal associated to an oscillatory process is an AS (ref. \cite{San18}). In particular, in \cite[Rueda et al.][]{Rue21a} the parametric expression for the AS  is given as: $\nu(t)=\sum_{J=1}^{m} A_J\sin(\phi_J(t)) $. Furthermore, in this paper, the AS is used to define the $Y$ axis in 2D and 3D representations of ECG signals  by considering that when $\mu^{II}(t)=P(\vec{D}(t)|LeadII)$, then $\nu^{II}(t)=P(\vec{D}(t)|LeadII^{(i)})$.
\\

 The 3DFMM$_{ecg}$ model relies on two premises. Namely;  1.-
the heart electric field originates from a multi dipole; and
2.- each dipole is represented by a vector, which moves in a plane as time progresses and  is mathematically described as a complex FMM  signal.

 
 Below, we prove that the projection of  such a vector is an FMM wave with the property that the values of $\alpha$ and $\omega$ do not depend on the direction of the projection. 
\\

\subsubsection*{Theorem}

Let \{$\vec{d}(t), t \in (0,2\pi] $\} be a set of vectors in the same plane, $ \wp $, and assume that the projection of  $\vec{d}(t)$ in a given direction $L$, such as $L \in \wp$, is an FMM wave, as follows;

\begin{equation}
P(\vec{d}(t)|L)=M^L+W(t_,A^L,\alpha^L,\beta^L,\omega^L), t \in (0,2\pi] \nonumber
\end{equation}

Then, the projection of $\vec{d}(t)$ in any other direction, $L'$, is also an FMM wave, verifying that $\alpha^{L'}=\alpha^L$ and $\omega^{L'}=\omega^L$.
\\
The proof of this theorem is given in the Supplementary Information.
\\

Now, consider the tridimensional space with its origin of the central point in the chest and the voltage recorder with the ten electrodes generating the  12-lead ECG signals: $Lset=\{I,II,III,aVR,aVL,aVF,V1,V2,V3,V4,V5,V6\}$. Each of these signals is the projection of $\vec{D}(t)$ in a given direction and, in turn, the projection of $\vec{D}(t)$ is the sum of the projections of the five dipole vectors, for each of which the result of the theorem can be applied individually.

Finally, the 3DFMM$_{ecg}$ model is derived, taking into account all the above considerations. Specifically, the ECG unidimensional signals  are formulated as a signal plus error model in Definition \ref{eq:3DFMM}, where the error term accounts for artefacts in the data and the intercept accounts for location changes as follows.

\begin{Def}{ \textit{ 3DFMM$_{ecg}$ model}} \label{eq:3DFMM}{\hspace{200mm}}	
Let $X^L(t_i),$ $t_1<...< t_n $,  be an observation from lead $L,$ $L \in  Lset$. Then,

		\begin{equation}\label{eq:mod}
	X ^L(t_i)= M^L+ \sum_{J \in \{P,Q,R,S,T\}} W(t_i,A_J^L,\alpha_J,\beta_J^L,\omega_J)+ e^L(t_i);
	\end{equation}

	\quad

 where, for  $ L \in Lset$, and, $J\in \{P,Q,R,S,T\}$:
			\begin{itemize}
			\item $ M^L \in \Re $,
			\item $\beta_J^L \in  (0,2\pi]$,
			\item $\alpha_P \le \alpha_Q \le \alpha_R \le \alpha_S \le \alpha_T \le \alpha_P$,
			\item $\omega_J \in  [0,1]$,
			\item $A_J^L \in \Re^+ $,
			
		\item $(e^L(t_1),...,e^L(t_n))' \sim N_n(\boldsymbol{0},\sigma^L\boldsymbol{I})$.
	\end{itemize}
	\vspace{1mm}
\end{Def}

It is relevant to note that restrictions imposed on the $\alpha$ parameters, response to the assumption that the atrial depolarization is previous to the ventricles depolarization. Occasionally,  the atria may repolarize later. In such a case, the $P$ wave would not go before the $QRS$. To prevent these cases, the model can also be defined by numbering the waves and making a subsequent identification of the numbers with the letters, maintaining the relationship between the $QRS$ and the $T$.

 Other important parameters of practical use are the peak and trough times. They are defined as functions of the basic parameters \cite{Rue21b}. In addition,  other indices, such as the distances between waves, are easily derived from the set of basic 3DFMM$_{ecg}$ model parameters. Obtaining good estimators of these indices is crucial for clinicians  because they are  useful tools in the diagnosis.

\subsection*{Evaluation metrics}
In this study, a global model quality measure, named $\overline{R}$, is defined for a given patient. Namely,
$$ \overline{R}= \frac{1}{12} \sum_{L \in Lset} median(R^2_{L_b}), $$
where, for a fixed beat $b$ and lead $L$,
$$R^2_{L_b}=1-\frac{\sum_{i=1}^n(X_b^L(t_i)-\hat{X_b^L}(t_i))^2}{\sum_{i=1}^n(X_b^L(t_i)-\overline{X}_b^L(t_i))^2}$$ 
 is the proportion of variation in the signal that is explained by the predicted values. The higher the $\overline{R}$, the better the model.
\subsection*{Identification algorithm}
 The identification problem reduces to solving the following optimization problem:
\begin{equation*}\label{eq:opt}
Min_{ \theta \in \Theta} \sum_{L \in Lset} \frac{1}{\sigma^L} \sum_{i=1}^n [ X^L(t_i)-\mu^L(t_i,\theta)]^2,
\end{equation*}

where $\theta$ is the vector of the model parameters, $\Theta$ is the parametric space, and $\mu^L(t_i,\theta)= M^L+ \sum_{J \in \{P,Q,R,S,T\}} W(t_i,A_J^L,\alpha_J,\beta_J^L,\omega_J) $.
 For a typical ECG pattern, $\Theta$ is initially defined as in Definition \ref{eq:3DFMM}. However,
 $\Theta$ is reduced when atypical or very noisy patterns  are analysed to achieve a correct physiological identification of waves. The values $\sigma^L, L\in Lset$ are identified as part of the optimization process. 
 
For computational efficiency, the $Lset$ is reduced to $Lred=\{I,II,V1,V2,V3,V4,V5,V6\}$, as the rest of the lead signals are linear combinations of $I$ and $II$. However, the weight of $I$ and $II$ increase by a factor of 3 in order to maintain the global weight of the derivations from the frontal plane\cite{Bay07}.

From a statistical point of view, the optimization problem defined above is that of finding the Maximun Likelihood Estimates of the parameters. We adapt an iterative algorithm\cite{Rue21a}  to the multivariate setting. The algorithm alternates  M and I steps.
 It is  assumed that the preprocessing stage provides a $QRS$ annotation point, denoted as $t^{QRS}$, for each beat. Moreover, exceptionally, one or more than one, of the eight leads are discarded in the preprocessing stage due to noise artefacts; only information on at least one of $I$, $II$, $V2$, or $V5$ lead is required.  In those cases, the identification stage is conducted with the selected leads, providing estimates for the common parameters; while the estimates for the lead-specific parameters of missing leads are derived by solving a standard multiple linear regression problem at the end of each M step.

 The M step obtains $K \ge 5$  FMM waves using a backfitting algorithm  and  the I step assigns $K \le 5$ letters to, at most, five of these waves. Typically, $K=5$;  however, in the presence of significant noise or when the morphology is pathological, it is possible that the interesting waves may be hidden between the sixth and seventh waves (exceptionally up to the tenth). The values $\sigma^L$ are updated in each iteration of the algorithm, averaging the squared difference between the expected values and the predicted values by the model. They are initially set to 1.

\textbf{M step}:  A standard backfitting algorithm is designed by fitting a single FMM wave simultaneously to the leads in \textit{Lred}.  The fitting of a single  FMM is repeated successively to the residuals. The numbers of backfitting passes programmed in each step M is five. 
The final estimates of $M^L$, $A_J^L$ and $\beta_J^L$ ; $L \in Lred ; J\in \{P,Q,R,S,T\},$ are derived by solving a standard multiple linear regression problem.  
\\

\textbf{I  step}: The $R$ wave is assigned in the first I step. It corresponds to the one with the highest explained variability among those closest to $t^{QRS}$, which also has a positive peak on leads $I$ or $II$ and a negative peak in lead $V2$. In the few cases where these assumptions do not hold, additional conditions are used.
Next, the preassignation of $P, Q, S$ and $T$ to the free components among the first five is done using $\alpha_P \le\alpha_Q \le \alpha_R\le \alpha_S \le \alpha_T$.  This preassigment corresponds, in most cases, to the  final assignment. 
However, in the presence of significant noise,  or when the morphology is pathological, the interesting waves may  sometimes be hidden between the sixth and seventh waves, which are checked for  reassignments  using thresholds on the main model parameters. In particular, noisy components are detected with too small or too high  $\omega$ values.

The algorithm finishes when there is no significant increase in the objective function of the optimization problem.
\\

In addition, the predictions from other leads in the frontal plane, $III$, $aVL$, $aVR$ and $aVF$ are derived  using the predictions of  the $I$ and $II$ leads, using the known linear relations between them \cite{Bay07}. In the case that any of the $I$ and $II$ leads is tentatively eliminated in the preprocessing step,  the estimates  for leads in the frontal plane can be  derived by solving a standard multiple linear regression problem, as  is done with the missing leads in step M. A flow chart of the algorithm is given in the Extended Data Fig. \ref{f:alg}.

\subsection*{Data and Code availability}

This work makes use of the PTB-XL database \cite{Wag20b} that is  publicly available at Physionet \cite{Gol00}.

All the presented results are reproducible through the 3DFMM$_{ecg}$ app, publicly available at \url{https://fmmmodel.shinyapps.io/fmmEcg3D/}, as it implements the API used for the analyses.  
A detailed description of the  3DFMM$_{ecg}$ app,  including  implementation and usage examples, is provided in the Supplementary Information. The corresponding author will provide the source code  upon reasonable request.


\bibliography{referenciasECG3D.bib}

\section*{Acknowledgements}
The authors gratefully acknowledge the financial support received by
the Spanish Ministry of Science, Innovation and Universities  [PID2019-106363RB-I00  to C.R., I.F. and Y.L.].

\section*{Author contributions}
C.R. conceived the aims, theoretical proposal, mathematical model developments and wrote the manuscript. C.R. and A.R.-C. developed the design and implementation of the identification algorithm.   C.R., Y.L. and  I.F. designed the preprocessing stage and the simulations.  A.R.-C. and C.C. developed the computational code and the app. M.D.U provided critical feedback on the manuscript. All the authors revised and approved the manuscript. 
\section*{Competing Interests}
The authors declare no competing interests.

\section*{Additional information}
\subsection*{Supplementary Information}
This file contains a  data preprocessing overview, the Theorem's proof,  a simulation study and a description of the 3DFMM$_{ecg}$ app.

\subsection*{Materials \& Correspondence}

Correspondence and requests for materials  should be addressed to	Dr. Rueda. email: \url{cristina.rueda@uva.es}

\newpage
\section*{Extended Data Figures and Tables}

\begin{suppfigure}[H]
\centering
\includegraphics[width=1\textwidth]{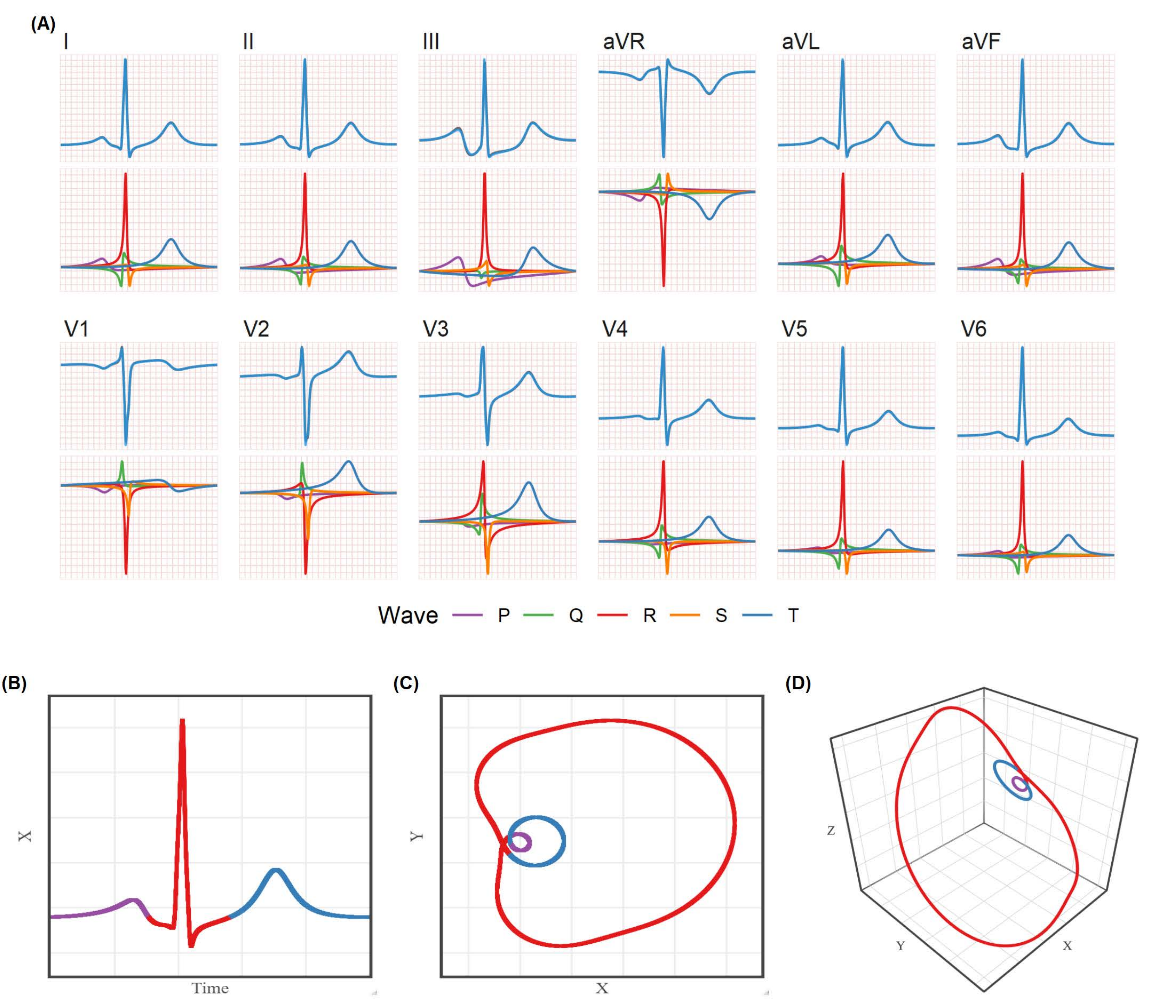}
\caption{\label{fig:simNORM}ECG signal from a NORM heartbeat generated by the 3DFMM$_{ecg}$ model. (A) 12-lead signals. (B), (C), (D): 1D, 2D and 3D representation of the predicted signal, respectively.}
\end{suppfigure}

\begin{suppfigure}[H]
\centering
\includegraphics[width=1\textwidth]{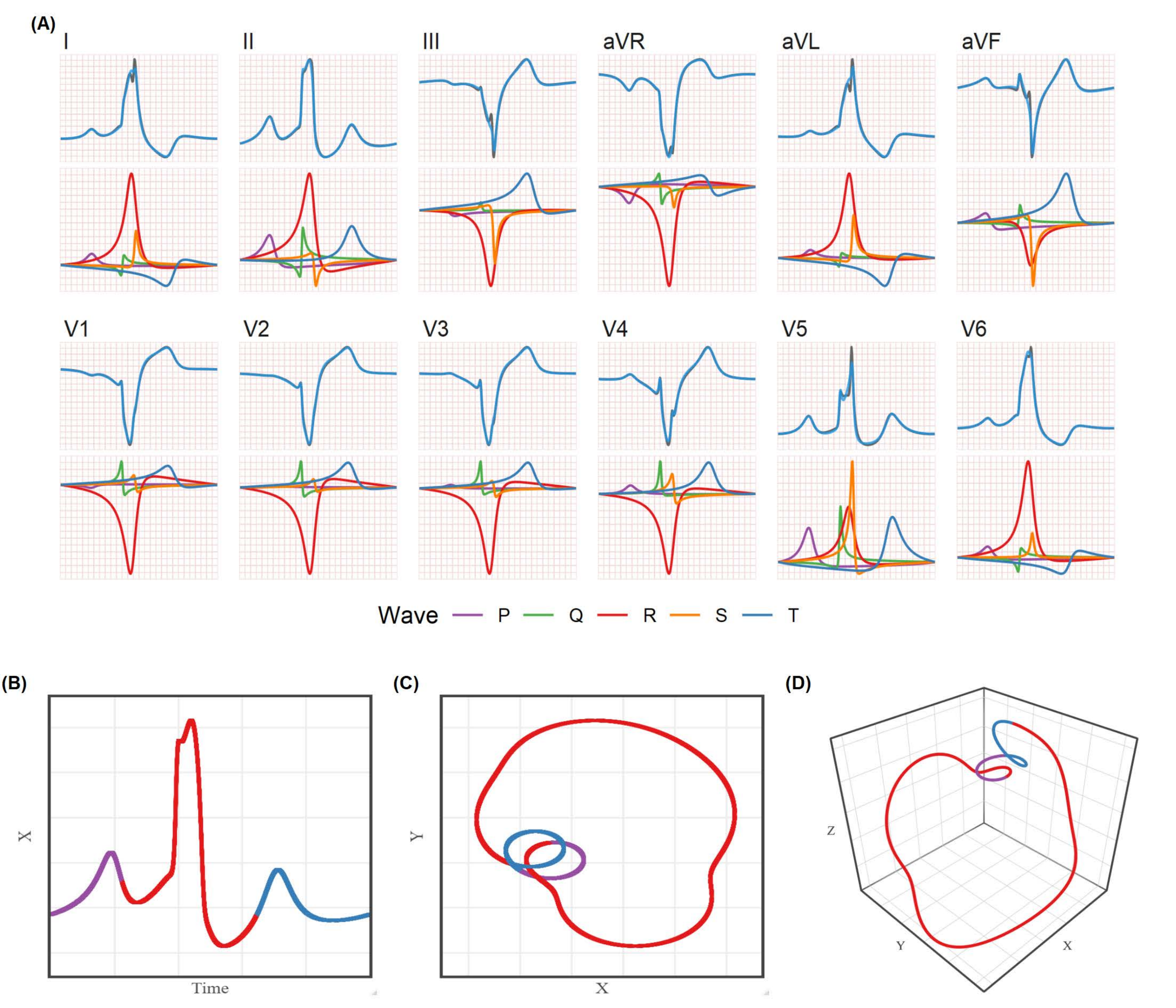}
\caption{\label{fig:simCLBBB}ECG signal from a CLBBB heartbeat generated by the 3DFMM$_{ecg}$ model. (A) 12-lead signals. (B), (C), (D): 1D, 2D and 3D representation of the predicted signal, respectively.}
\end{suppfigure}

\begin{suppfigure}[H]
\centering
\includegraphics[width=1\textwidth]{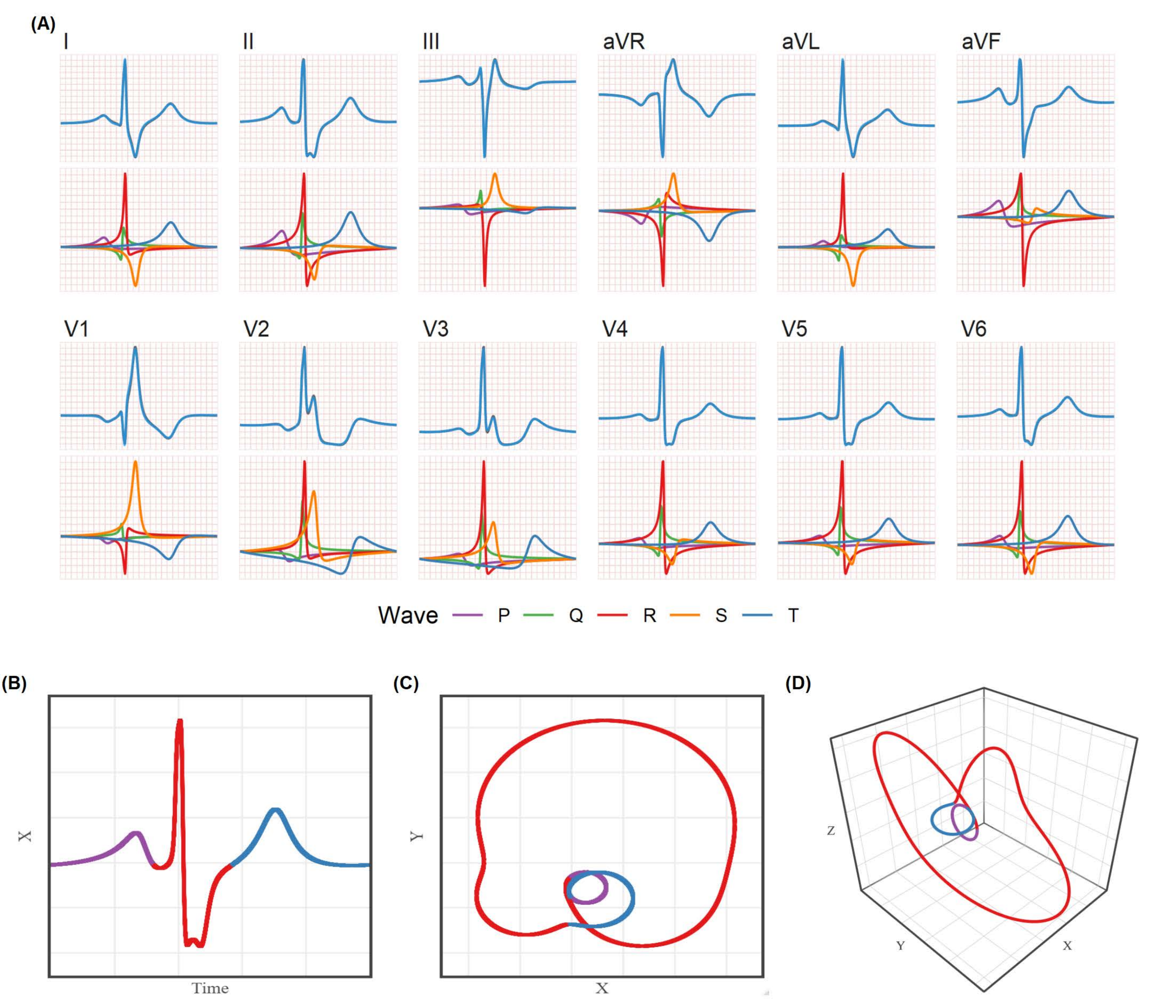}
\caption{\label{fig:simCRBBB}ECG signal from a CRBBB heartbeat generated by the 3DFMM$_{ecg}$ model. (A) 12-lead signals. (B), (C), (D): 1D, 2D and 3D representation of the predicted signal, respectively.}
\end{suppfigure}

\begin{suppfigure}[H]
\centering
\includegraphics[width=1\textwidth]{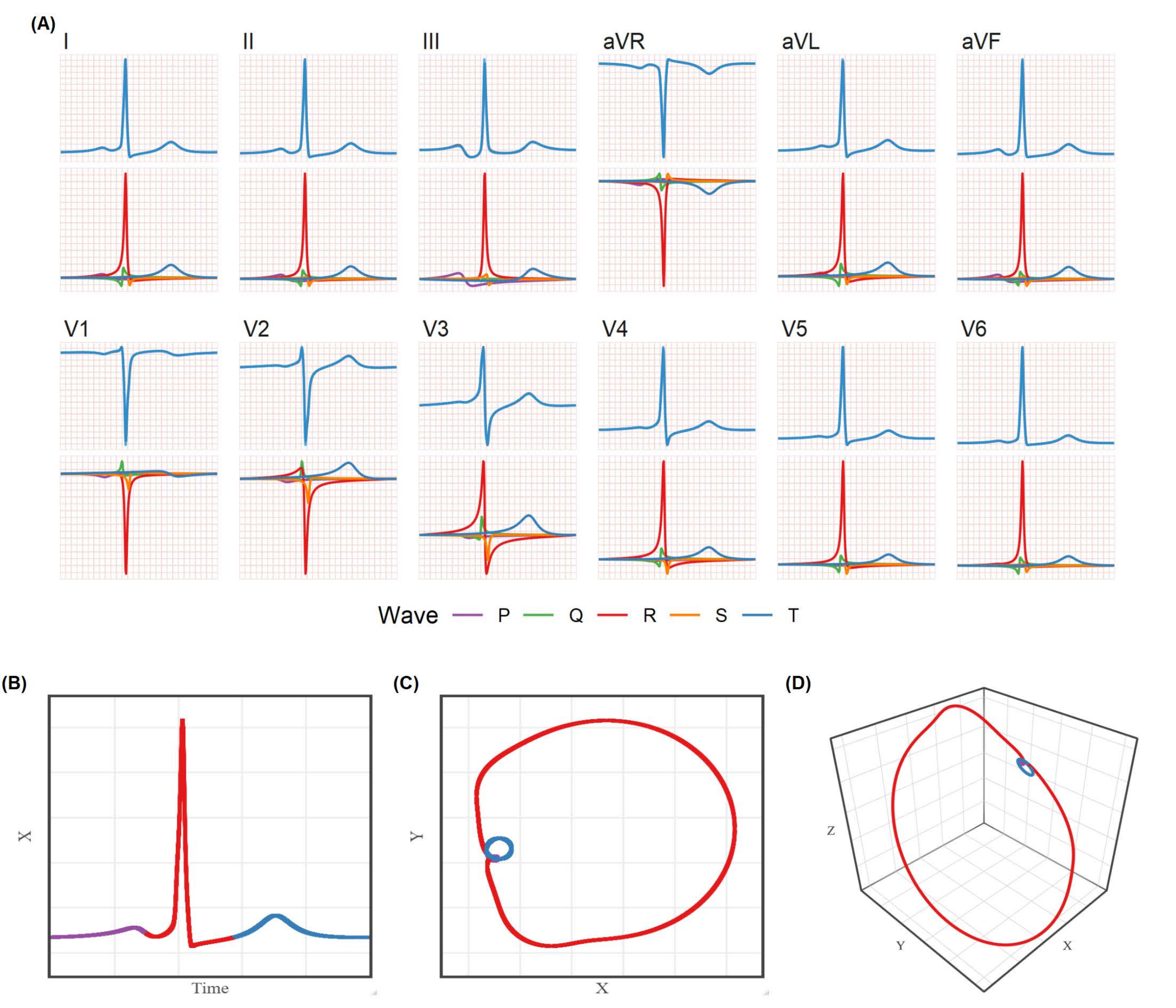}
\caption{\label{fig:simLVH}ECG signal from a HYP heartbeat generated by the 3DFMM$_{ecg}$ model. (A) 12-lead signals. (B), (C), (D): 1D, 2D and 3D representation of the predicted signal, respectively.}
\end{suppfigure}

\begin{suppfigure} [H]
\centering
\includegraphics[width=1\textwidth]{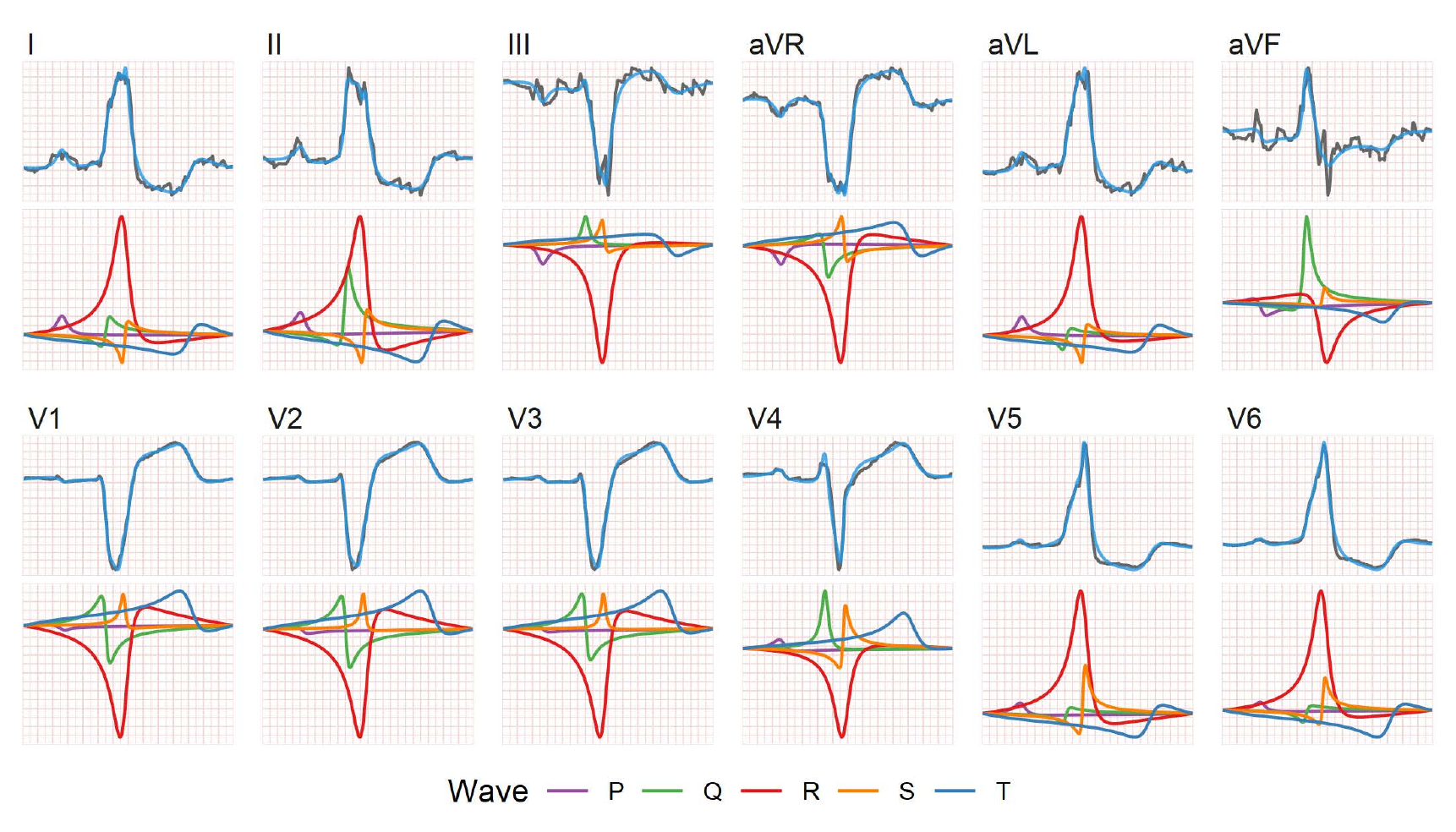}
\caption{\label{fig:Ex1}12-lead ECG signal from patient id $1710$ in PTB-XL database, beat nº3. Cardiological diagnostic: CLBBB.}
\end{suppfigure}

\begin{suppfigure} [H]
\centering
\includegraphics[width=1\textwidth]{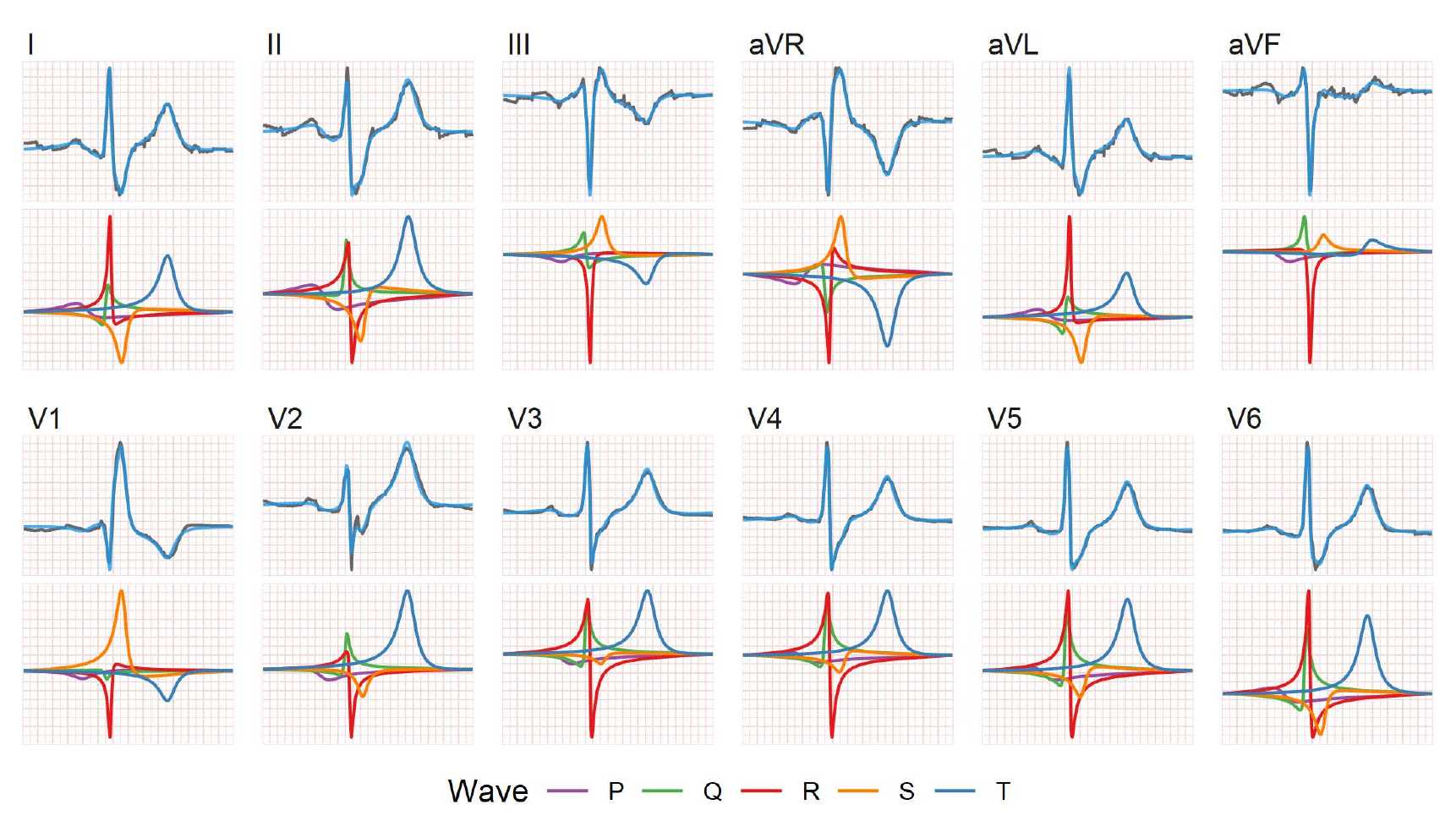}
\caption{\label{fig:Ex2}12-lead ECG signal from patient id $195$ in PTB-XL database, beat nº 1. Cardiological diagnostic: CRBBB.}
\end{suppfigure}

\begin{suppfigure} [H]
\centering
\includegraphics[width=1\textwidth]{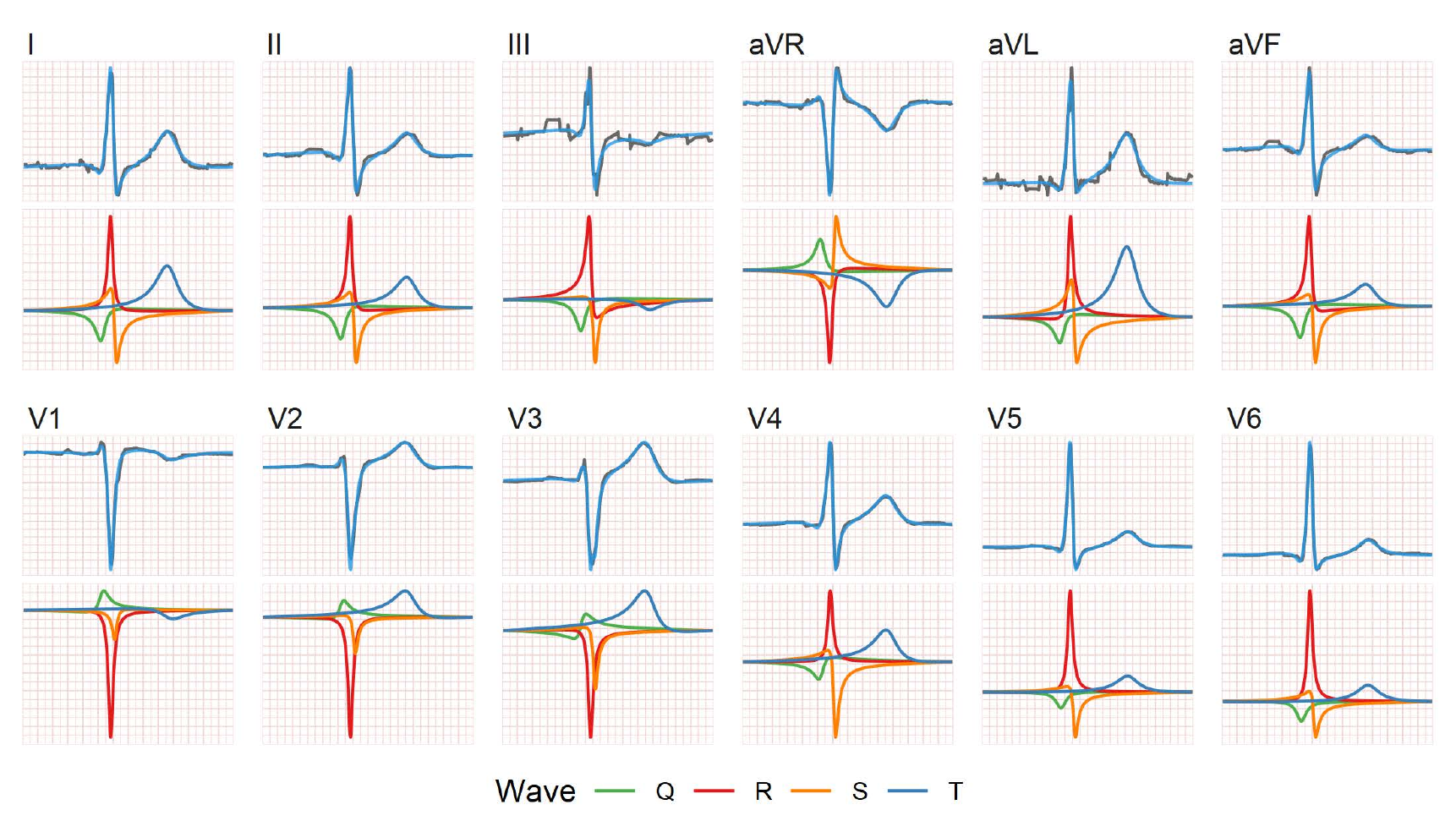}
\caption{\label{fig:Ex3}12-lead ECG signal from patient id $30$ in PTB-XL database, beat nº6. Cardiological diagnostic: HYP.}
\end{suppfigure}

\begin{suppfigure} [H]
\centering
\includegraphics[width=1\textwidth]{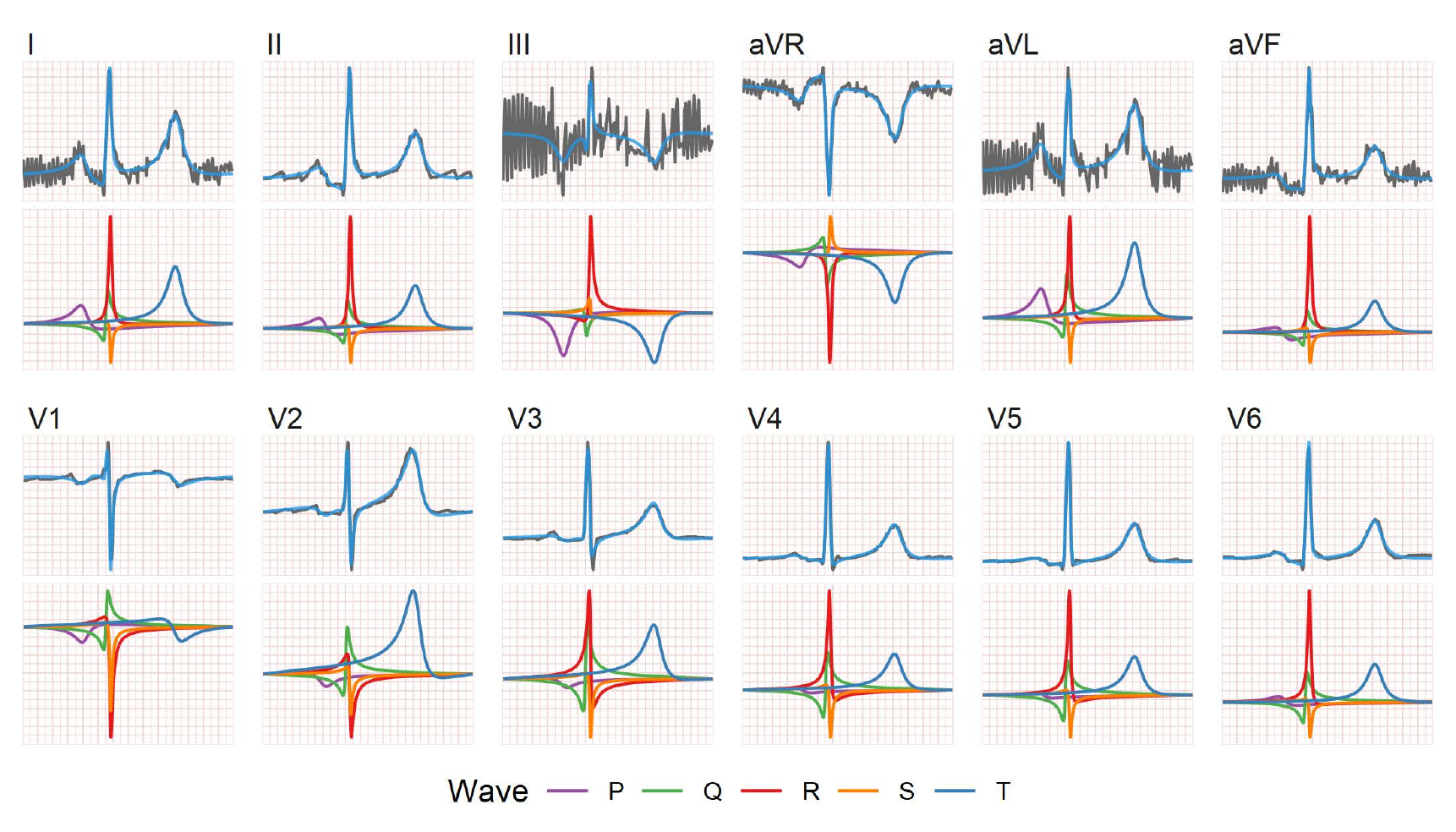}
\caption{\label{fig:Ex4}12-lead ECG signal from patient id $2838$ in PTB-XL database, beat nº7. Noisy ECG signal for a NORM subject.}
\end{suppfigure}

\begin{suppfigure} [H]
\centering
\includegraphics[width=1\textwidth]{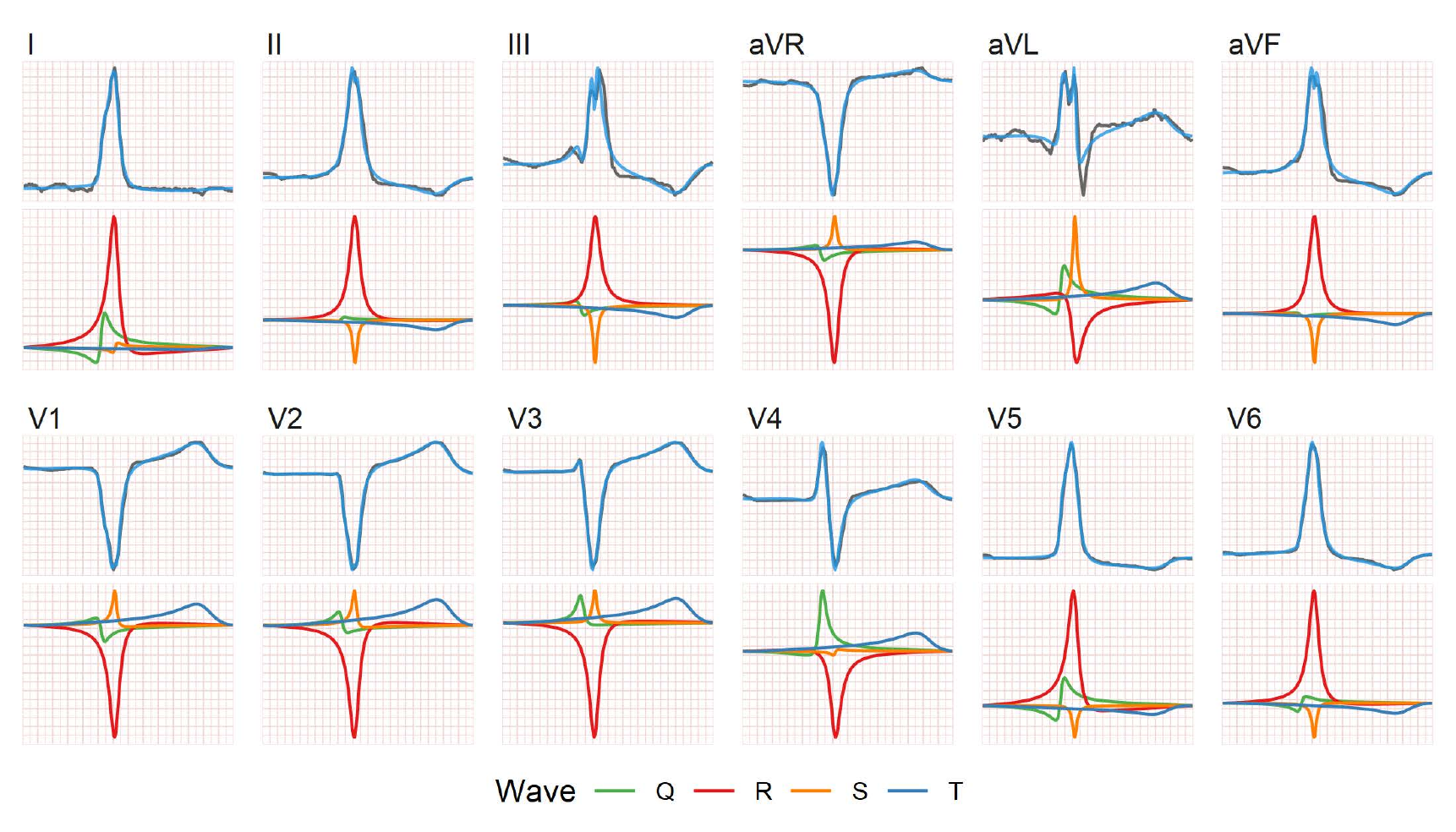}
\caption{\label{fig:Ex5}12-lead ECG signal from patient id $12507$ in PTB-XL database, beat nº11. Cardiological diagnostic: Incomplete LBBB. Absence of $P$ wave.}
\end{suppfigure}


\begin{suppfigure}[H]
\includegraphics[width=1\textwidth]{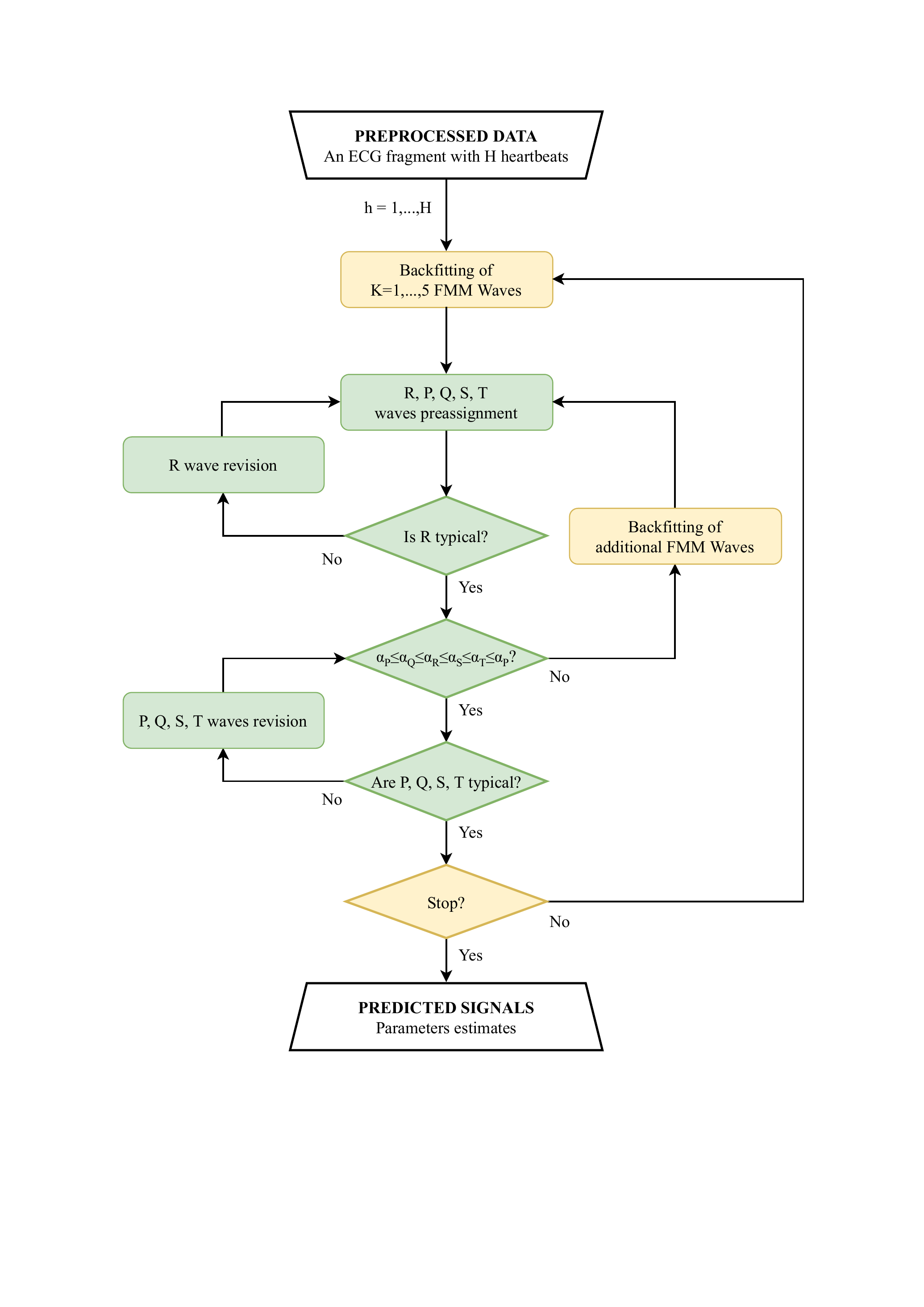}
      \caption{Flow chart of the Identification Algorithm. The yellow and green blocks correspond to the  M-step and the I-step, respectively. White blocks refer to the inputs and outputs. }
          \label{f:alg}
\end{suppfigure}

\end{document}